\documentclass{ws-p10x7}

\newcommand{\beq}{\begin{equation}}
\newcommand{\eeq}{\end{equation}}
\newcommand{\ba}{\begin{array}}
\newcommand{\ea}{\end{array}} 
\newcommand{\beqa}{\begin{eqnarray}}
\newcommand{\eeqa}{\end{eqnarray}}

\newcommand{\cL}{{\cal L}}

\newcommand{\cA}{{\cal A}}
\newcommand{\cO}{{\cal O}}

\newcommand{\Br}{{\cal B}}

\newcommand{\no}{\nonumber}
\newcommand{\lsim}{\stackrel{<}{_\sim}}

\newcommand{\tu}{{\tilde u}}

\newcommand{\ttop}{{\tilde t}}

\newcommand{\cne}{C_9^{\rm eff}}

\newcommand{\ct}{C_{10}}

\def\npb#1#2#3{    {\it Nucl. Phys. }{\bf B #1}, #3 (#2)}
\def\plb#1#2#3{    {\it Phys. Lett. }{\bf B #1}, #3 (#2)}
\def\prd#1#2#3{    {\it Phys. Rev. }{\bf D #1}, #3 (#2)}

\def\prl#1#2#3{    {\it Phys. Rev. Lett. }{\bf #1}, #3 (#2)}
\def\ptp#1#2#3{    {\it Prog. Theor. Phys. }{\bf #1}, #3 (#2)}

\def\rmp#1#2#3{    {\it Rev. Mod. Phys. }{\bf #1}, #3 (#2)}

\def\jhep#1#2#3{   {\it JHEP  }{\bf #1}, #3 (#2)}

\begin{document}

\title{$Z$ penguins and rare B decays}

\author{Gino Isidori}

\address{INFN, Laboratori Nazionali di Frascati, 
                I-00044 Frascati, Italy \\E-mail: isidori@lnf.infn.it}
%
%
%

\twocolumn[\maketitle\abstract{
Rare $B$ decays of the type  $b \to s~\ell^+\ell^-(\nu\bar\nu)$
are analyzed in a generic scenario where New Physics effects 
enter predominantly via $Z$ penguin contributions. We show that  
this possibility is both phenomenologically allowed and 
well motivated on theoretical grounds.
The important role played in this context
by the lepton forward-backward 
asymmetry in $B\to K^*\ell^+\ell^-$ is emphasized.
}]

\section{Introduction}

Flavour-changing neutral-current (FCNC) processes provide a powerful tool in 
searching for clues about non-standard flavour dynamics. Being generated only at the 
quantum level and being additionally suppressed, within the Standard Model (SM), 
by the smallness of the off-diagonal entries of the 
Cabibbo-Kobayashi-Maskawa (CKM) matrix,\cite{CKM} 
their observation is very challenging. This suppression, however, 
ensures a large sensitivity to possible non-standard effects, 
even if these occur at very high energy scales,
rendering their experimental search highly valuable.

In the present talk we focus on a specific class of non-standard 
$\Delta B=1$ FCNC transitions: those mediated by
the $Z$-boson exchange and contributing to rare $B$ decays 
of the type $b \to s~\ell^+\ell^-(\nu\bar\nu)$.
As we shall show, these 
are particularly interesting for two main reasons:
i) there are no stringent experimental bounds 
   on these transitions yet;
ii) it is quite natural to conceive extensions of 
the SM where the $Z$-mediated FCNC amplitudes
are substantially modified, even taking into 
account the present constraints on $\Delta B=2$
and $b \to s\gamma$ processes.

In a generic extension of the Standard Model where new 
particles appear only above some high scale $M_X > M_Z$, 
we can integrate out the new degrees of freedom and generate 
a series of local FCNC operators already at the electroweak scale.
Those relevant for $b \to s~\ell^+\ell^- (\nu\bar{\nu})$
transitions can be divided into three wide classes: 
generic dimension-six operators, magnetic penguins and
FCNC couplings of the $Z$ boson.\cite{BHI}
The latter are dimension-four operators of the type 
$\bar{b}_{L(R)} \gamma^\mu s_{L(R)} Z_\mu$, that we are allowed to
consider due to the spontaneous breaking of $SU(2)_L \times U(1)_Y$.
Their coefficients must be proportional to some symmetry-breaking 
term but do not need to contain any explicit $1/M_X$ suppression
for dimensional reasons,
contrary to the case of dimension-six operators and magnetic penguins.
This naive argument seems to suggest that 
FCNC couplings of the $Z$ boson are  particularly 
interesting and worth to be studied independently of the other effects.
It should be noticed that the requirement of naturalness in the size of 
the $SU(2)_L \times U(1)_Y$ breaking terms implies that also 
the adimensional couplings of the 
non-standard $Z$-mediated FCNC amplitudes must vanish
in the limit $M_X/M_Z \to \infty$. Nonetheless, as we will illustrate 
below with an explicit example,
the above naive dimensional argument remains a strong indication 
of an independent behaviour of these couplings with respect 
to the other FCNC amplitudes. 

\section{FCNC $Z$ penguins in generic SUSY models}
An explicit example where the largest deviations from the 
SM, in the sector of FCNC, are generated by the $Z$ boson exchange
can be realized within supersymmetric models 
with generic flavour couplings. Within this context,
assuming $R$ parity conservation and  
minimal particle content, FCNC amplitudes 
involving external quark fields turn out to 
be generated only at the quantum level.
Moreover, assuming the natural 
link between trilinear soft-breaking terms 
and Yukawa couplings, sizable $SU(2)_L$- and flavour-breaking 
effects can be expected in the up sector due 
to the large Yukawa coupling of the third generation.
Thus the potentially dominant non-SM effects
in the effective $Z\bar{b}s$ vertex turn out to be 
generated by chargino-up-squarks loops 
and have a pure left-handed structure, 
like in the SM.\cite{LMSS}

Similarly to the $Z\bar{s}d$ case,\cite{CI} 
the first non-vanishing contribution appears 
to the second order in a simultaneous expansion of
chargino and squark mass matrices in the basis of 
electroweak eigenstates. The potentially 
largest effect arises when the necessary $SU(2)_L$ breaking 
($\Delta I_W =1$) is equally shared by the
$ \ttop_R -\tu^{s}_L $ mixing 
and by the chargino-higgsino mixing, 
carrying both $\Delta I_W =1/2$.
For a numerical evaluation, 
normalizing the SUSY result to the SM one 
(evaluated in the 't~Hooft-Feynman gauge) and varying 
the parameters in the allowed ranges, leads to:\cite{BHI,LMSS}
\beqa
\left| \frac{ Z_{sb}^{\rm SUSY} }{ Z_{sb}^{\rm SM} } \right|   
&\lsim&  \frac{ 0.1}{|V_{ts}|} \left| \frac{( M^2_{\tilde U})_{t_R s_L} }{  M^2_{\tu_L} } \right|  
 \left( {M_W \over M_2} \right) \no \\ 
 &=& 2.5 \left| (\delta^U_{RL})_{32}  \right| \left( {M_W \over M_2} \right)~.
\label{eq:Zsusynum}
\eeqa
The coupling $(\delta^U_{RL})_{32}$, which represents the analog of the 
CKM factor $V_{ts}$ in the SM case, is not very constrained 
at present and can be of $\cO(1)$
with an arbitrary $CP$-violating phase.
Note, however, that vacuum stability bounds\cite{Casas}  
imply  $| (\delta^U_{RL})_{32} | \lsim \sqrt{3} m_t/M_S$, where 
$M_S$ denotes the generic scale of sparticle masses. 
Therefore the SUSY contribution to the $Z$ penguin decouples
as $(M_Z/M_S)^2$ in the limit $M_S/M_Z \to \infty$.

As it can be checked by the detailed analysis of 
Lunghi {\it et al.},\cite{LMSS} 
in the interesting scenario where the left-right 
mixing of up-type squarks is the only non-standard 
source of flavour mixing, $Z$ penguins 
are largely dominant with respect to other supersymmetric 
contributions to $b\to s~\ell^+\ell^-$.
Indeed, due to the different $SU(2)_L$ structure,
the $\ttop_R -\tu^{s}_L $ mixing contributes to magnetic penguins 
only to the third order in the mass expansion discussed above.
Therefore in this scenario the magnetic-penguin contribution to  
$b\to s~\ell^+\ell^-$ is additionally suppressed by $M_Z/M_S$
with respect to the $Z$-penguin one.
Similarly, in the case of box diagrams the $\ttop_R -\tu^{s}_L $ mixing
alone leads to a contribution that decouples like $M^4_Z/M^4_S$.
%
%

\section{Experimental bounds on the $Z\bar{b}s$ vertex}
An extended discussion of other non-standard scenarios 
where large deviations form the SM occur in the 
$Z\bar{b}s$ vertex can be found elsewhere.\cite{BHI} 
We proceed here analyzing the experimental information 
on this FCNC amplitude in a model-independent way.

The dimension-four effective FCNC couplings of the $Z$ boson
relevant for  $b\to s$ transitions can be
described by means of the following effective Lagrangian
\beqa
  \label{eq:Zsb}
  && \cL^{Z}_{FC} = \frac{G_F}{\sqrt{2}} \frac{e}{ \pi^2} M_Z^2
  \frac{\cos \Theta_W}{\sin \Theta_W} Z^\mu  \no\\ 
&&  \times \left( 
   Z^L_{sb}~\bar b_L \gamma_\mu s_L 
+ Z^R_{sb}~\bar b_R \gamma_\mu s_R \right) \,+\, {\rm h.c.}, \quad\
\eeqa
where $Z^{L,R}_{sb}$ are complex couplings.
Evaluated in the  't~Hooft-Feynman gauge,  
the SM contribution to $Z^{L,R}_{sb}$ is given by 
\beq
  \label{eq:SMZsb} 
  Z^R_{sb}\vert_{\rm SM} = 0~,\ \
  Z^L_{sb}\vert_{\rm SM} = V_{tb}^* V_{ts} C_0(x_t)~,
\eeq
where $x_t=m_t^2/m_W^2$ and $C_0(x)$ is a loop 
function\cite{IL,BBL} of $O(1)$. 
Although $Z^{L}_{sb}\vert_{\rm SM}$ is not gauge invariant, we recall 
that the leading  contribution to both $b \to s~\ell^+\ell^- $ and  
$b \to s~\nu\bar\nu $ amplitudes in the limit $x_t \to \infty$ 
is gauge independent and is generated by the 
large $x_t$ limit of $Z^{L}_{sb}\vert_{\rm SM}$
($C_0(x_t) \to x_t/8$ for $x_t\to\infty$). 

Constraints on $|Z^{L,R}_{sb}|$ can be obtained from the experimental 
upper bounds on exclusive and inclusive  $b \to s~\ell^+\ell^- (\nu\bar\nu)$ 
transitions. The latter are certainly more clean form the theoretical
point of view (especially the $b \to s~\nu\bar\nu$ one\cite{GLN}) although 
their experimental determination is quite difficult. At present the 
most significant information from exclusive decays is given by\cite{CLEO}
$\Br(B\to X_s \ell^+ \ell^- ) < 4.2 \times 10^{-5}$
and leads to\cite{BHI}
\beq
\left( \left| Z^L_{sb} \right|^2 + \left| Z^R_{sb} \right|^2 \right)^{1/2}
   \lsim 0.15~.
\label{eq:Zsblim}
\eeq
Within exclusive channels the most stringent information 
can be extracted from $B\to K^* \mu^+\mu^-$, where the 
experimental upper bound\cite{CDF} on the non-resonant branching 
ratio ($\Br^{\rm n.r.} <  4.0 \times 10^{-6}$) lies only about 
a factor two above the SM expectation.\cite{ABHH99}
Taking into account the uncertainties on the hadronic form 
factors, this implies\cite{BHI}
\beq
 \label{eq:ZsblimCDFLR}
   \left| Z^{L,R}_{bs} \right| \lsim 0.13~.
\eeq

Additional constraints on the $ Z^{L,R}_{bs}$ couplings 
could in principle be obtained by the direct limits  
on $\Br(Z\to b \bar s)$ and by  $B_s- \bar B_s$ mixing,
but in both cases these are not very significant.

Interestingly the bounds (\ref{eq:Zsblim}-\ref{eq:ZsblimCDFLR})
leave open the possibility of large deviations from 
the SM expectation in (\ref{eq:SMZsb}). 
In the optimistic case where $Z^L_{bs}$
or $Z^R_{bs}$ were close to saturate these bound, we 
would be able to detect the presence of non-standard
dynamics already by observing sizable rate enhancements 
in the exclusive modes. In processes like 
$B\to K^* \ell^+ \ell^-$ and $B\to K \ell^+ \ell^-$,
where the standard photon-penguin diagrams provide a 
large contribution, the enhancement could be at most of a factor 2-3. 
On the other hand, in processes like 
$B\to K^* \nu \bar{\nu}$,  $B\to K \nu \bar{\nu}$
and $B_s \to \ell^+ \ell^-$, where the photon-exchange 
amplitude is forbidden, the maximal enhancement could reach a 
factor 10. 

\section{Forward-backward asymmetry in 
$B\to K^*\mu^+\mu^-$}
If the new physics effects do not produce sizable 
deviations in the magnitude of the  
$b \to Z^* s$ transition, it will be hard to 
detect them from rate measurements, 
especially in exclusive channels. 
A much more interesting observable in this respect 
is provided by the forward-backward (FB) asymmetry of the
emitted leptons, also within exclusive modes.
In the  $\bar B\to \bar K^* \mu^+\mu^-$ case this is defined as 
\beqa
&& \cA^{(\bar B)}_{FB}(s)=\frac{1}{d\Gamma(\bar B\to \bar K^* \mu^+\mu^- )/ds}
  \int_{-1}^1 \!\!\! d\cos\theta 
\no\\
&&  
\label{eq:asdef} \qquad
\frac{d^2 \Gamma(\bar B\to \bar K^* \mu^+\mu^- )}{d s~ d\cos\theta}
\mbox{sgn}(\cos\theta)~,
\eeqa
where $ s =m_{\mu^+\mu^-}^2/m_B^2$ and 
$\theta$ is the angle between the momenta of 
$\mu^+$ and $\bar B$ in the dilepton center-of-mass frame. 
Assuming that the leptonic current has only a 
vector ($V$) or axial-vector ($A$) structure, then the
FB asymmetry provides a direct measure of 
the $A$-$V$ interference. 
Since the vector current is largely dominated by 
the photon-exchange amplitude and the axial one is 
very sensitive to the $Z$ exchange, 
$\cA^{(\bar B)}_{FB}$ and $\cA^{(B)}_{FB}$
provide an excellent tool
to probe the  $Z\bar{b}s$ vertex.

Employing the usual notations for the Wilson coefficients of the 
SM effective Hamiltonian relevant to $b\to s~\ell^+\ell^-$
transitions,\cite{BBL} $\cA^{(\bar B)}_{FB}(s)$ turns out to be 
proportional to\footnote{~To simplify the notations we have 
introduced the parameter $\cne(s)$ that is not a Wilson 
coefficient but it can be identified with $C_9$ at the 
leading-log level.\cite{BHI}}
\beq
  {\rm Re}\left\{  \ct^* \left[ s~\cne(s) 
    + \alpha_+(s) \frac{m_b C_7}{m_B}  \right] \right\}, 
  \label{eq:dfbabvllex}
\eeq
where  $\alpha_+(s)$ is an appropriate ratio 
of hadronic form factors.\cite{BHI,burdman0}
The overall factor ruling the magnitude of $\cA^{(\bar B)}_{FB}(s)$
is affected by sizable theoretical 
uncertainties. Nonetheless there are at least 
three features of this observable
that provide a clear short-distance information:

i) Within the SM $\cA^{(\bar B)}_{FB}(s)$ has a zero
in the  low $s$ region ($s_0|_{\rm SM} \sim 0.1$).\cite{burdman0}
The exact position of $s_0$ is not free from 
hadronic uncertainties at the $10\%$ level, 
nonetheless the existence of the zero itself is 
a clear test of the relative sign between 
$C_7$ and $C_9$. The position of $s_0$ is 
essentially unaffected by possible
new physics effects in the  $Z\bar{b}s$ vertex.

\begin{figure}
\centerline{\epsfysize=2.5in{\epsffile{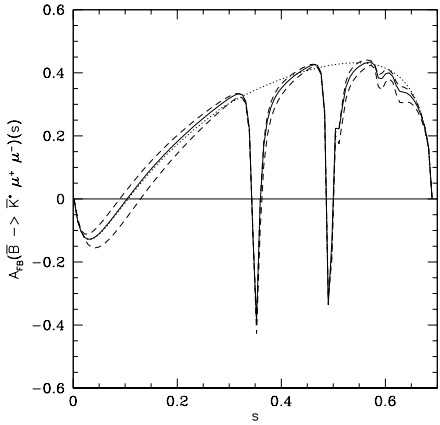}}}
\caption{ \it FB asymmetry of
$\bar B\to \bar K^* \mu^+\mu^-$ within the SM.
The solid (dotted) curves have been
obtained employing the Krueger-Sehgal\protect\cite{ks96} approach 
(using the perturbative end-point effective 
Hamiltonian\protect\cite{BHI}). 
The dashed lines show the effect of varying the
renormalization scale of the Wilson Coefficients
between $m_b/2$ and $2 m_b$,
within the Krueger-Sehgal approach. }
\label{fig:AFB}
\end{figure}

ii) The sign of $\cA^{(\bar B)}_{FB}(s)$ around the zero
is fixed unambiguously in terms of the relative sign
of $C_{10}$ and $C_9$:\cite{BHI} within the SM one 
expects $\cA^{(\bar B)}_{FB}(s) > 0$ for $s>s_0$,
as in Fig.~\ref{fig:AFB}.
This prediction is based on a model-independent 
relation among the form factors\cite{LEET} 
that has been overlooked in most of the recent 
literature.  Interestingly, the sign of $C_{10}$
could change in presence of a non-standard   
$Z\bar{b}s$ vertex leading to a striking signal
of new physics in $\cA^{(\bar B)}_{FB}(s)$, 
even if the rate of $\bar B \to \bar K^* \ell^+\ell^-$
was close to its SM value.

iii) In the limit of $CP$ conservation one expects 
$\cA^{(\bar B)}_{FB}(s) = - \cA^{(B)}_{FB}(s)$.
This holds at the per-mille level within the 
SM, where $C_{10}$ has a negligible $CP$-violating phase,
but again it could be different in presence 
of new physics in the  $Z\bar{b}s$ vertex.
In this case
the ratio $[ \cA^{(\bar B)}_{FB}(s) + \cA^{(B)}_{FB}(s)]/
[ \cA^{(\bar B)}_{FB}(s) - \cA^{(B)}_{FB}(s)]$
could be different from zero, for $s$ above the charm threshold, 
reaching the $10\%$ level in realistic models.\cite{BHI}

\section*{Acknowledgements}
I am grateful to G. Buchalla and G. Hiller for the
enjoyable collaboration on this subject.


\end{document}